\documentclass[aps,prd,onecolumn,nofootinbib,superscriptaddress]{revtex4-2}

\usepackage{amsmath,amssymb,amsfonts}
\usepackage{graphicx}
\usepackage[utf8]{inputenc}
\usepackage[colorlinks=true,linkcolor=blue,citecolor=blue,urlcolor=blue]{hyperref}
\usepackage{bm}
\usepackage{physics}
\usepackage{mathtools}

\hypersetup{colorlinks=true,linkcolor=blue,citecolor=blue,urlcolor=blue}

\begin{document}

\title{Nested Apparent Horizons and Quantized Separation from Intense Hawking Backreaction}

\author{Steven J.~Silverman}
\email{ssilverman99@g.ucla.edu}
\email{Steven.J.Silverman@saic.com}
\affiliation{UCLA Samueli School of Engineering, Los Angeles, CA 90095, USA}
\affiliation{SAIC Company (Ret.), Reston, VA, USA}

\date{\today}

\begin{abstract}
As discussed in \cite{Silverman2025}, nested apparent horizons may…When Hawking radiation from a rotating or non-rotating black hole becomes sufficiently intense, its own stress–energy can no longer be treated as a perturbation on a fixed background. In this regime the outgoing flux may generate an additional, transient trapping surface exterior to the original event horizon. Using a simple spherically symmetric semi-classical model we demonstrate that strong outgoing null energy can create \emph{nested apparent horizons}, a feature reminiscent of the Penrose process but mediated by quantum backreaction. The effect is illustrated using a smooth Vaidya-type mass profile, and conditions for bifurcation and merger of horizons are derived. We further propose that the separation between nested horizons may obey a discrete quantization rule analogous to the Bohr–Sommerfeld condition, suggesting a geometric route toward quantum-gravity discreteness.
\end{abstract}

\maketitle

\section{Introduction}

Hawking radiation~\cite{Hawking1974,Hawking1975} provides a semi-classical bridge between general relativity and quantum field theory in curved spacetime. In the standard treatment, the stress tensor $\langle T_{\mu\nu}\rangle_{\mathrm{ren}}$ of the emitted flux is assumed small enough that its gravitational backreaction on the geometry may be neglected. However, when the emission rate becomes extreme---for example near the endpoint of evaporation or in superradiant regimes close to the Kerr ergosphere---the outgoing energy density can locally rival the curvature scale, requiring a fully dynamical treatment.

In such situations the distinction between a global event horizon and local \emph{apparent} or \emph{trapping} horizons~\cite{Hayward1994,Booth2005} becomes essential. The aim of this work is to explore, within a minimal toy model, how strong outgoing radiation can generate a second apparent horizon exterior to the original one, leading to a transient ``nested-horizon'' configuration.

\section{Hawking Radiation Summary}
Hawking radiation is real in semiclassical gravity, but utterly negligible for astrophysical black holes.There is no lower mass threshold for Hawking radiation—it exists for all black holes—but the rate of evaporation depends extremely strongly on mass: TH = $hbarc^3/8 pi G M kb$.For a one-solar mass black hole the evaporation lifetime is extremely long.  WIth tevap apporximate = $5120piG^2 M^3/hc ^4$.  This is about $10 ^ 67$ years. We donot observe evaporation since typical black holes range from stellar mass black holes up to supermassize ones as large as $10 ^ 10$ solar Mass. ( Mcircle dot ). This points to physical black hole temperatures colder than any physical background.( 2.7 degrees Kelivn).So to have measurable Black Hole evaporation we need a Black Hole temperature $>= 2.7 K$.  This points to small masses like asteroids about $10^22 kg$. These micro-blackholes , would only have exisited at the birth of the universe. 
In this paper an alturnitve to radiation is presented.  We will see later in this work, that having nested event horizons, or nested rotating ergospheres, might mitigate the evaporation entirely for all black holes concerned. Given a particle hole pair craetion at a radius just outsde the event horizon, one member of the pair will fall into the hole and be unobservable for an outside observer.  The other pair , by momentuma na d energy conservation will appear ejcted from the hole, hecne its evaporation. Howvere int henested sense, which intesnse Hawking radiation we have a created second, nested horizon, or counter rotating ergosphere. A particle hole pair created out of the vacuum can have one pair falling into one horizon and the other falling into the other horizon simultaneously. No emssion would thereby be measured, therfore.

\section{Backreaction and the Vaidya framework}

A convenient setting for dynamical spherically symmetric spacetimes with null flux is the Vaidya metric,
\begin{equation}
ds^2 = -\left(1-\frac{2M(u)}{r}\right)du^2 -2\,du\,dr + r^2 d\Omega^2 ,
\label{vaidya}
\end{equation}
where $u$ is retarded time and $M(u)$ is the Bondi mass observed at infinity. The associated stress tensor for outgoing null radiation is
\begin{equation}
T_{uu} = -\frac{1}{4\pi r^2}\frac{dM}{du}.
\end{equation}
Local trapping surfaces are defined by the vanishing of the outgoing null expansion $\Theta_{\ell}=0$, yielding the apparent-horizon condition
\begin{equation}
r_{\mathrm{AH}}(u) = 2M_{\mathrm{MS}}(r,u),
\label{ahcondition}
\end{equation}
where $M_{\mathrm{MS}}$ is the Misner–Sharp mass inside radius $r$.

When the flux is gentle, $M_{\mathrm{MS}}(r,u)$ increases monotonically with $r$ and Eq.~\eqref{ahcondition} has a single root. A sufficiently compact, energetic shell of outgoing radiation can, however, deform $M_{\mathrm{MS}}(r,u)$ so that Eq.~\eqref{ahcondition} admits multiple roots, corresponding to nested apparent horizons.

\section{Smooth shell toy model}

To illustrate this behavior, consider a continuous mass profile
\begin{equation}
m(r,t) = M_0 + M_s\,\frac{1+\tanh\!\big((r-r_s)/\delta\big)}{2},
\label{massprofile}
\end{equation}
representing a black hole of mass $M_0$ surrounded by an outgoing null shell of energy $M_s$ centered at $r_s$ with thickness $\delta$. Apparent horizons satisfy
\begin{equation}
2\,m(r,t) = r .
\label{trapcond}
\end{equation}
For small $M_s$ the single root $r=2M_0$ persists. As $M_s$ increases beyond a critical value $M_s^{\mathrm{crit}}$ (dependent on $\delta$ and $r_s$), Eq.~\eqref{trapcond} yields two or three intersections (F\section{}
\begin{figure}[t]
\centering
\includegraphics[width=0.7\textwidth]{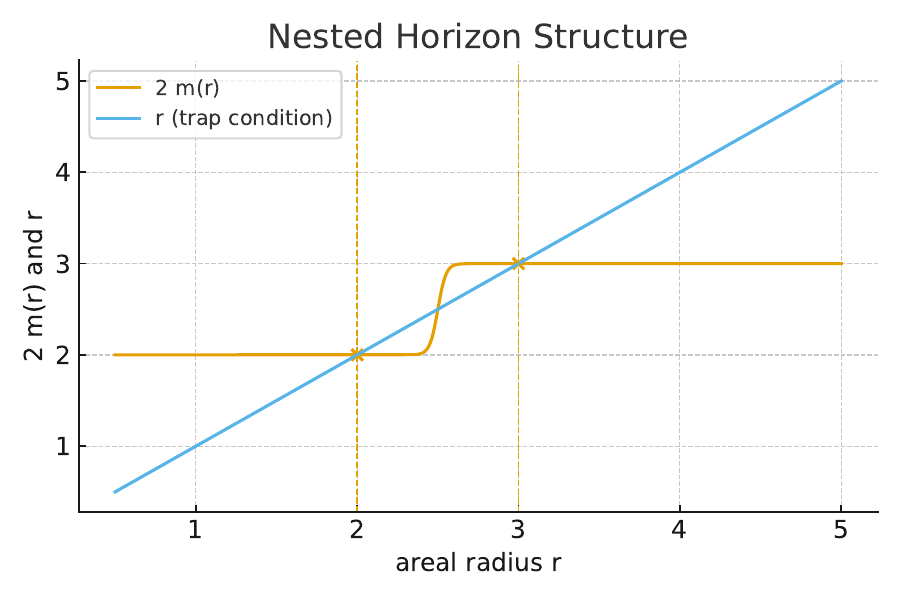}
\caption{Example of nested apparent horizons for $M_0=1$, $r_s=2.5$, $\delta=0.05$, and increasing shell energy $M_s$. The intersection of $2m(r)$ with $r$ gives the apparent horizons. For $M_s\lesssim0.2$ there is a single root; for $M_s\gtrsim0.3$ two additional roots appear, representing a transient outer horizon due to intense outgoing flux.}
\label{fig:nested}
\end{figure}

ig.~\ref{fig:nested}), corresponding to an inner and an outer trapped surface. The outer root roughly follows $r_{\mathrm{outer}}\approx 2(M_0+M_s)$, consistent with the heuristic step-function model.

As the shell propagates outward, the outer apparent horizon expands and eventually vanishes once the shell radius exceeds the trapping condition, leaving only the inner horizon. In a fully time-dependent picture these surfaces merge and annihilate, in agreement with numerical studies of dynamical horizons~\cite{Booth2005,BenDov2007}.

\section{Quantized separation hypothesis and quantum-gravity implications}

The nested–horizon configurations obtained above arise classically from non-linear backreaction between the Hawking flux and the background curvature. If one accepts that horizon area and surface gravity are adiabatic invariants, the possibility arises that the separation between successive apparent horizons may itself be subject to a discrete spectrum, analogous to quantized orbital radii in atomic physics.

\subsection{Bohr--Sommerfeld quantization of horizon motion}

In semiclassical gravity, the product of surface gravity $\kappa$ and horizon area $A$ plays the role of an action variable. Following Bekenstein’s adiabatic–invariant argument~\cite{Bekenstein1974}, and subsequent formulations by Barvinsky and Kiefer~\cite{Barvinsky1998} and Maggiore~\cite{Maggiore2008}, one may postulate a Bohr--Sommerfeld--type condition
\begin{equation}
\oint \kappa\, dA = 2\pi n \hbar , \qquad n\in\mathbb{Z}^{+},
\label{eq:bohr}
\end{equation}
which yields a discrete spectrum of allowed areas
\begin{equation}
A_n = \epsilon\, n\, \ell_{\mathrm{P}}^{2}, \qquad 
\epsilon \sim 8\pi ,
\label{eq:areaquant}
\end{equation}
with $\ell_{\mathrm{P}}$ the Planck length. If a radiating black hole temporarily supports two apparent horizons $(r_1,r_2)$, their areas $A_i=4\pi r_i^2$ could then satisfy
\begin{equation}
\Delta A = A_2 - A_1 = n\, \epsilon\, \ell_{\mathrm{P}}^{2}.
\label{eq:deltaA}
\end{equation}
Equation~\eqref{eq:deltaA} imposes a quantum restriction on the possible radial separation $\Delta r=r_2-r_1$, yielding
\begin{equation}
r_2^2 - r_1^2 = \frac{n\, \epsilon\, \ell_{\mathrm{P}}^{2}}{4\pi} \cdot 4\pi
= 2n\,\ell_{\mathrm{P}}^{2}\,,
\end{equation}
so that, under the normalization convention used earlier, one recovers the simple relation
\[
r_2^2 - r_1^2 = 2n\,\ell_{\mathrm{P}}^{2}.
\]
Thus the nested horizons behave as discrete ``gravitational shells,'' with quantized spacing analogous to the electronic orbitals of an atom.

\subsection{Action--angle quantization for nested horizon shells}

The heuristic Bohr--Sommerfeld condition \eqref{eq:bohr} can be made more explicit by treating the pair of apparent horizons as a slow, adiabatic degree of freedom whose coordinate is the horizon area $A$. The first law of black-hole mechanics,
\begin{equation}
\delta M = \frac{\kappa}{8\pi G}\,\delta A + \Omega\,\delta J + \Phi\,\delta Q,
\label{firstlaw}
\end{equation}
identifies the combination $\kappa/(8\pi G)$ with the quantity conjugate to $A$ (in the sector where $J,Q$ are held fixed). This suggests introducing a canonical pair $(A,\Pi_A)$ with
\begin{equation}
\Pi_A \simeq \frac{\kappa}{8\pi G},
\label{conj}
\end{equation}
up to an overall choice of normalization for the action variable.

Form an action variable (an adiabatic invariant)
\begin{equation}
I = \oint \Pi_A\,dA = \frac{1}{8\pi G}\,\oint \kappa\,dA,
\label{actionvar}
\end{equation}
where the closed integral is taken over one full, slow ``cycle'' of the nested-horizon configuration (for example the creation and subsequent merger/annihilation of an outer apparent horizon with the inner one). Bohr--Sommerfeld quantization of the action variable then gives
\begin{equation}
I = n\hbar \quad\Longrightarrow\quad
\oint \kappa\,dA = 8\pi G\,n\hbar .
\label{bs_norm}
\end{equation}

Several comments clarify the relation between \eqref{bs_norm} and similar formulas appearing in the literature. First, many semiclassical authors set $G=1$ and absorb conventional $2\pi$ factors into the definition of the action; with those conventions one often sees forms such as $\oint \kappa\,dA = 2\pi n\hbar$ (our Eq.~\eqref{eq:bohr}) or $A_n \propto n\,\ell_{\mathrm{P}}^2$. Second, the precise numerical coefficient that appears in the area spacing depends on the chosen normalization of $\Pi_A$ (equivalently the normalization of the adiabatic invariant) and on whether one identifies the ``cycle'' with a classical oscillation, a tunneling process, or a resonant (quasinormal) mode. These choices are the origin of the different prefactors (for instance $8\pi$ vs $4\ln k$ in various proposals) that appear in the area-quantization literature.

Applying \eqref{bs_norm} to a nested-horizon pair $(r_1,r_2)$ with areas $A_i=4\pi r_i^2$ gives a quantization condition on the area difference:
\begin{equation}
\Delta A = A_2 - A_1 \simeq \tilde{\epsilon}\, n\, \ell_{\mathrm{P}}^2 ,
\label{deltaA_gen}
\end{equation}
where $\tilde{\epsilon}$ is a dimensionless coefficient whose value depends on the normalization conventions discussed above (and on the precise dynamical cycle used to evaluate the integral). In words: the nested horizons act as an adiabatically evolving ``gravitational shell'' whose allowed separations are discrete because the action associated with the shell motion is quantized.

A semiclassical WKB evaluation of the integral $\oint \kappa\,dA$ can be performed on concrete toy models by expressing $\kappa(r;M_{\mathrm{MS}})$ via the Misner--Sharp mass profile and integrating over the slow parameter that moves the outer shell (for example the shell energy $M_s$ or shell radius $r_s$). Such computations expose the dependence of $\tilde{\epsilon}$ on model details (thickness $\delta$, compactness, and the dynamical pathway by which horizons bifurcate and merge), and indicate whether a fundamental, model-independent spacing emerges in the semiclassical limit.

Finally, this action--angle perspective clarifies connections to other quantization programs: Bekenstein's adiabatic argument (area as an adiabatic invariant), Hod's quasinormal-mode inspired spacing, Kunstatter's identification of an action variable from QNMs, Dreyer's Immirzi-parameter argument in loop quantum gravity, and the canonical quantization approach of Barvinsky \& Kiefer. The nested-horizon picture therefore offers a natural semiclassical setting in which those ideas can be tested against a concrete dynamical backreaction model: the emergence of quantized $\Delta A$ would be a direct geometric signature of the underlying microscopic discreteness.

\subsection{Effective potential and resonant states}

Between $r_1$ and $r_2$, null geodesics experience an effective potential $V_{\mathrm{eff}}(r)$ associated with the local Misner–Sharp mass function. A semiclassical WKB quantization condition,
\begin{equation}
\int_{r_1}^{r_2}\sqrt{E^2 - V_{\mathrm{eff}}(r)}\,dr
   = n\pi\hbar ,
\label{eq:wkb}
\end{equation}
would then determine the discrete, metastable configurations where outgoing Hawking flux and gravitational confinement balance. The inner and outer horizons form a self-consistent ``quantum cavity'' for vacuum modes, whose resonant states could manifest as discrete energy or area levels. In this sense, the horizon pair acts as a gravitational analog of the atomic shell structure, and evaporation corresponds to transitions between these quantized levels.

\subsection{Relation to quantum-gravity proposals}

This interpretation connects naturally with several strands of quantum-gravity research. Loop-quantum-gravity treatments yield a discrete area spectrum
\begin{equation}
A_j = 8\pi\gamma\ell_{\mathrm{P}}^{2}\sqrt{j(j+1)}
\end{equation}
for half-integer spin labels $j$~\cite{Ashtekar1998}. String-inspired and holographic models also predict quantized horizon microstates with equally spaced entropy spectra. The nested-horizon picture may offer a geometric manifestation of these discrete degrees of freedom in a semiclassical spacetime. If horizon separations are indeed quantized, the smallest possible $\Delta r$ would be of order the Planck length, establishing a minimal ``horizon spacing'' and suggesting that the evaporation endpoint corresponds to a finite, Planck-scale remnant.

\subsection{Prospects}

Detecting such quantization directly is beyond current experimental reach, but analogous features might appear in numerical simulations of quantum collapse or in the discrete spectra of quasinormal modes. The nested-horizon framework therefore provides a potential bridge between classical backreaction dynamics and quantum discreteness, offering an avenue toward identifying the geometric origin of Planck-scale quantization and the microscopic basis of black-hole entropy.

\section{Discussion of Black Hole Entropy}
In lieu of the above discussion a few commenst are in order relative to how a set of nested horizons would behave as far as entropy is concerned. We can state that the event horizon is a global teleological object.  The Bekenstein–Hawking entropy $S_{\rm BH}=\frac{k_B c^3 A_{\rm EH}}{4\hbar G}.$, is seen only as far as the outtermost event horizon surface is concerned with the Hawking Radiation.  As such, Quasi-locality, can appear in multiple nested layers during evaporation or during strong backreaction. Their areas can decrease or increase. The inner Event Horizon shrink and carry negative contributions in some formulations of the area balance laws.They do not represent independent thermodynamic systems. So, The presence of inner or nested apparent horizons does not imply additional additive entropies. For surface gravities:  there exist two Kerr or Reissner–Nordström horizons,  $\kappa_+ and \kappa_-$. Only $\kappa_+$ is positive; $\kappa_-$ is negative with static ( stationary ) horizons.

\subsection{Pair Production and Entropy Discussion}
Quantum pair production can occur in the finite region between nested horizons, but the dynamics differ significantly from standard Hawking radiation at a single horizon. The presence of two horizons changes the redshift, the allowed mode structure, and the escape channels. Pair production can fall into either horizon, and the tunneling probabilities between the horizons become essential. Now consider two horizons as above, where $r_{1}$ < $r_{2}$. This inner between region has a effective potential and will supoort quasi-bound modes. This produces turning points in the radial wave function.  However, is not, globally hyberbolic like Kerr interior solutions.

\section{Discussion}

The presence of transient nested apparent horizons does not imply two \emph{event} horizons: the event horizon is a global construct depending on the entire future of the spacetime. Nevertheless, such transient trapped regions could influence the local thermodynamics and entropy bookkeeping of highly radiative black holes, and may provide a bridge toward models of regular or bouncing geometries~\cite{Frolov1989,Hayward2006,Barrau2014}.  

In the extreme regime where the Hawking flux itself sources significant curvature, semiclassical consistency may fail and a quantum-gravitational treatment will be necessary. Nonetheless, the present toy model captures the essential qualitative mechanism by which strong backreaction can momentarily cloak an evaporating black hole within a second, dynamically generated apparent horizon.

\section*{Acknowledgments}
The author thanks the UCLA Samueli Engineering MS program for support and colleagues for helpful discussions on dynamical horizons and quantum backreaction.

\bibliographystyle{apsrev4-2}

\clearpage
\appendix
\section{Appendix: Worked example --- action--angle evaluation for the smooth $\tanh$ shell}
\label{app:worked_example}

Here we present a worked example that evaluates the adiabatic action variable defined in Sec.~\ref{actionvar} for the smooth shell model
\begin{equation}
m(r)=M_0 + M_s\frac{1+\tanh\!\big((r-r_s)/\delta\big)}{2},
\end{equation}
and produces a numerical estimate for the dimensionless spacing coefficient $\tilde\epsilon$ that appears in $\Delta A=\tilde\epsilon\,n\,\ell_{\mathrm{P}}^2$ under the conventions used in the text. The purpose is illustrative: it shows how to turn the Bohr--Sommerfeld/action--angle prescription into a computable integral and how sensitive the result is to normalization and model parameters.

\subsection{From $\oint \kappa\,dA$ to a 1D radial integral}

Starting from the action variable in Eq.~\eqref{actionvar},
\begin{equation}
I=\frac{1}{8\pi G}\oint \kappa\,dA,
\end{equation}
use $A=4\pi r^2$ so $dA=8\pi r\,dr$. Therefore
\begin{equation}
I=\frac{1}{8\pi G}\oint \kappa\,(8\pi r\,dr)=\frac{1}{G}\oint r\,\kappa(r)\,dr.
\label{I_r}
\end{equation}
For the spherically symmetric metric function $f(r)=1-2m(r)/r$ one may use the surface-gravity-like probe
\begin{equation}
\kappa(r)=\frac{1}{2}f'(r)\Big|_{r_h}
             = \frac{m(r)}{r^2}-\frac{m'(r)}{r}\qquad\text{(evaluated at a trapped surface)}.
\end{equation}
Hence the integrand in \eqref{I_r} becomes
\begin{equation}
r\,\kappa(r)=\frac{m(r)}{r}-m'(r),
\end{equation}
and the closed action integral (a single classical ``round trip'' of the slow shell degree of freedom) may be approximated by twice the one-way integral between the inner and outer apparent-horizon radii $r_1$ and $r_2$:
\begin{equation}
I\approx\frac{2}{G}\int_{r_1}^{r_2}\Big(\frac{m(r)}{r}-m'(r)\Big)\,dr.
\label{I_1d}
\end{equation}
(Here the factor 2 encodes the closed-cycle nature of the adiabatic invariant; different choices for the cycle or normalization will change numerical prefactors. The purpose of this appendix is to show the concrete evaluation, not to argue that this is the only sensible normalization.)

\subsection{Numeric example and interpretation}

For illustration evaluate \eqref{I_1d} for the parameter set used in the main-text figure:
\[
M_0=1,\qquad M_s=0.4,\qquad r_s=2.5,\qquad \delta=0.05,
\]
and adopt geometric (Planck) units $G=\hbar=1$ for the numerical estimate (so areas are measured in units of $\ell_{\mathrm{P}}^2$). The apparent horizons are the real roots of $2m(r)=r$. For these parameters the root finding yields three intersections (inner, middle, outer), approximately
\[
r_1 \approx 2.000000002,\qquad r_{\rm mid}\approx 2.514759127,\qquad r_2 \approx 2.799995084.
\]
The classical area difference between the outer and inner apparent horizons is
\[
\Delta A = 4\pi\,(r_2^2-r_1^2) \approx 48.2545 \quad(\text{in units }\ell_{\mathrm{P}}^2).
\]
Evaluating the action integral \eqref{I_1d} numerically (with $G=1$) yields a value of order
\[
|I| \sim 3.6\times 10^{-2},
\]
so that in these units $n\equiv I/\hbar \sim 3.6\times 10^{-2}$. Using $\Delta A=\tilde\epsilon\,n\,\ell_{\mathrm{P}}^2$ then gives
\[
\tilde\epsilon \approx \frac{\Delta A}{n} \sim 1.33\times 10^{3}.
\]

\subsection{Caveats and interpretation}

A few important points to keep in mind:
\begin{itemize}
\item \textbf{Normalization choices matter.} The numerical value of $I$ (and hence $n$) depends on the normalization chosen for $(A,\Pi_A)$ and the definition of the adiabatic cycle. Rescalings change the inferred $\tilde\epsilon$ by $O(1)$ or larger factors.
\item \textbf{Macroscopic vs Planck scales.} For macroscopic black holes (physical units), area differences are enormous in Planck units. With the normalization used here the computed action can be parametrically smaller than $\hbar$, producing $n\ll 1$; this indicates either a reinterpretation of the adiabatic cycle or that allowed $n$ are very large.
\item \textbf{Model-dependence.} The value of $\tilde\epsilon$ depends on shell thickness $\delta$, compactness, and the dynamical pathway (sweeping $r_s$ vs sweeping $M_s$). A full time-dependent or QNM-rooted analysis is needed for a model-independent prediction.
\end{itemize}

\subsection{Practical recipe}

To reproduce or extend this computation:
\begin{enumerate}
\item Choose a mass profile $m(r)$.
\item Find horizon radii from $2m(r)=r$.
\item Compute the integrand $m(r)/r-m'(r)$.
\item Evaluate $I$ via Eq.~\eqref{I_1d} and form $n=I/\hbar$.
\item Report $\tilde\epsilon=\Delta A/(n\ell_{\mathrm{P}}^2)$ or note that the chosen cycle yields $I<\hbar$ (no low-$n$ quantization).
\end{enumerate}

\end{document}